\documentstyle[psfig,12pt,aaspp4]{article} 

\begin{document}

\title{Observations of Cygnus X-1 during the two spectral states
with the Indian X-ray Astronomy Experiment (IXAE) }
\author{A. R. Rao, P. C. Agrawal, B. Paul, M. N. Vahia, and J. S. Yadav}
\affil{Tata Institute of Fundamental Research, Homi Bhabha Road, Mumbai 400 005, India}
\affil{e-mail: arrao@tifrvax.tifr.res.in (ARR), pagrawal@tifrvax.tifr.res.in (PCA),
bpaul@tifrvax.tifr.res.in (BP), vahia@tifrvax.tifr.res.in (MNV) and jsyadav@tifrvax.tifr.res.in (JSY)}
	\and
\author{T. M. K. Marar, S. Seetha and K. Kasturirangan}
\affil{ISRO Satellite Centre, Airport Road, Vimanpura P.O. Bangalore  560 017, India.}
\affil{e-mail: seetha@isac.ernet.in (SS)}

\begin{abstract}
We present the time variability characteristics of Cygnus X-1 in its
two spectral states. The observations were carried out using the 
Pointed Proportional Counters (PPC) on-board the Indian X-ray Astronomy
Experiment (IXAE). The details of the instrument characteristics, the
observation strategy, and the background modeling methods are described.
In the soft state of Cyg X-1, we confirm the general trend of the Power 
Density Spectrum (PDS) obtained using the Proportional Counter Array 
(PCA) on-board the RXTE satellite.
The hard state of the source just prior to the spectral transition 
was not observed by the PCA and we present the PDS obtained in this state.
We find that the low frequency end of the PDS is flatter than that
observed during the spectral transition. Additionally, we find that
there is one more component in the low frequency end of the PDS, which 
is independent of the spectral state of the source. The time variability
is also examined by taking the statistics of the occurrence of shots
and it is found that the shot duration and shot energy follow an exponential
distribution, with time constants significantly different in the
two spectral states. In the soft state of the source a shot with a
very large strength has been identified and it has exponential rise
and decay phases with time constants of 0.4 s.  We examine these
results in the light of the current models for 
accretion onto black holes.

\end{abstract}

\section{Introduction}
 
Cygnus X-1 is a well-known Galactic black hole candidate.
It shows two distinct spectral states, `hard'
and `soft', and it spends most of its time in the hard state. 
The soft X-ray spectral shape
changes drastically between these two states, whereas the hard X-ray
spectrum is described by a power law with photon index $-$1.5$\pm$0.2
at all times (Liang \& Nolan 1984). 

  Rapid and chaotic intensity variations over time scales of milliseconds
to several seconds have been seen from this source (Oda 1977; Liang \&
Nolan 1984) and such variations have been conventionally taken as one of
the indicators of the existence of black holes. 
The power density spectrum obtained in the hard state (Nolan et al. 1981;
Belloni \& Hasinger 1990a) is flat below a certain frequency (about 0.1 Hz)
and decreases above that value.
The observed variability in Cyg X-1 is conventionally 
explained as random shots and the properties of these shots have been
studied extensively (Weisskopf et al. 1978; Lochner et al. 1991;
Negoro et al. 1994). Time delays between X-rays in different
energy ranges have been detected by Miyamoto \& Kitamoto (1989)
and these delays manifest as double-peaked structure in the phase lag
for different time scales and they have been interpreted as
arising from clumps of matter having two preferred sizes.
Though there are attempts to reproduce the power density spectrum from
simple shot models (Belloni \& Hasinger 1990a), the observed shape of the
power density spectrum is more complex than a simple shot noise model 
(Belloni \& Hasinger 1990b).

There have been attempts to model the wide-band X-ray spectrum
of Cyg X-1 and interpret it in terms of  more refined accretion 
theories. Using the simultaneous X-ray and gamma-ray data on Cyg X-1
obtained using Ginga and OSSE, respectively, Gierlinski et al. (1997)
found that the energy spectrum requires Comptonization components along with
additional components. Chitnis et al. (1997) used data obtained from
EXOSAT, OSSE and balloon-borne detectors and concluded that two
component thermal-Compton models  are sufficient to explain the wide
band data and the results are explained in terms of the accretion disk
theory developed by Chakrabarti \&  Titarchuk (1995). The same data, 
however, can also be explained by a transition disk model (along with its
reflection) with steeply varying temperature profiles (Misra et al. 1997). 
These results point towards the existence
of  Comptonizing plasma along with a reflecting material. 

On the theoretical side, 
Chakrabarti \&  Titarchuk (1995) have taken a complete solution of viscous
transonic equations and demonstrated that the accretion disk has a highly
viscous Keplerian part which resides on the equatorial plane and a sub-Keplerian 
component which resides above and below it. The sub-Keplerian component can
form a standing shock wave 
(or, more generally, a centrifugal barrier supported dense region)
which heats up the disk to a high 
temperature. The wide-band X-ray spectrum of Cyg X-1 qualitatively
agrees with this model (Chitnis et al. 1997). Narayan \& Yi (1994) have 
examined advection dominated accretion flows in black holes 
and their model is also used to explain qualitatively 
the spectral behavior and spectral states of Cygnus X-1.

 There  are, however, no quantitative attempts to explain both the spectral
and temporal properties of Cyg X-1 in its two states. Molteni et al. (1996)
do give a qualitative picture of how quasi-periodic
oscillations (QPO) can occur in black hole candidates. 
In recent times there are renewed attempts to give a complete physical
picture of accretion onto black holes because of the theoretical work
by Chakrabarti and collaborators and Narayan and co-workers. The 
transition of Cyg X-1 to a soft state in 1996 and its observation by several
X-ray satellites currently in orbit have helped in providing valuable
information for a proper  understanding of black hole accretion.

Cyg X-1 entered a rarely observed soft state in 1996 May (Cui 1996;
Zhang et al. 1996). The source remained in this state and started 
to go back to its normal hard state in 1996 July. The 
complete light curves in X-rays and gamma-rays have been tracked by the
ASM on board the RXTE satellite and the BATSE on-board the CGRO satellite.
A clear anti correlation was seen between the soft and hard X-rays (Cui et al.
1997a). Detailed spectral and temporal studies of the source
were carried out periodically with the PCA on-board the RXTE (Cui et al. 1997a,b;
Belloni et al. 1996). It was found that the spectral softening is associated
with changes in the power density spectrum (PDS) and also that the average
delay of hard photons relative to soft photons increases when the source
makes a transition from the soft state to the hard state.  These results
show that  the spectral and temporal 
parameters like the hardness ratio, time lag, PDS shape etc. change
monotonically with time during the spectral transition, possibly
indicating the existence of 
a single underlying mechanism responsible for both the spectral and 
temporal changes.

 We have observed the source in both the states using the Pointed 
Proportional Counters (PPC) on-board the Indian X-ray Astronomy Experiment
(IXAE). The soft-state observations are made during a period when 
there are no PCA observations and we present, for the first time,
the timing observations carried out in the hard state just prior to the
transition of the state. A preliminary report of these observations is
given in Agrawal et al. (1996).

      The paper is organized as follows. In section 2 we describe 
the  PPC observations and the  details of the background modeling. 
The results obtained from a detailed temporal analysis of the 
source are given in the next section.
In section 4 we discuss the results in the light of the current understanding
of black hole accretion and a brief 
summary of the main results of this work is given in the last section.

\section{IXAE instrument details and observations}

The Indian X-ray Astronomy Experiment (IXAE) includes three identical 
Pointed Proportional Counters (PPCs) and one X-ray Sky Monitor. Each PPC
is a multi-cell proportional counter array and has an effective area of 
400 cm$^2$. The filling gas is 90\% argon and 10\% methane at
a pressure of 800 mm of Hg.  
There are 54 cells with a size of 11 mm $\times$ 11 mm 
arranged in 3 layers. The bottom layer and the end cells are joined together
to form a veto output for charged particle anti-coincidence. The remaining 
anode cells in the top two layers form the detection volume and they operate
in mutual anti-coincidence. A passive collimator restricts the 
field of view to 2.3$^{\circ} \times$ 2.3$^{\circ}$. 
The operating energy range is between 2 keV and
18 keV. The overall energy resolution is 22\% at  6 keV. 
The gain stability of the detectors  is monitored continuously by X-rays 
from a collimated Cd$^{109}$ radioactive source irradiating the veto cells. 

 Each PPC has its own front-end electronics (consisting of amplifiers and
command-controllable high voltage unit) and a processing electronics. The
processing electronics selects the genuine events based on the pre-determined
logic conditions and measures the pulse height spectrum in 64 linear 
channels. Parallelly, independent counters store the following data i)
2 keV - 6 keV genuine events of top layer, ii) 2 keV - 18 keV genuine events 
of top layer, iii) 2 keV - 18 keV genuine events of middle layer, iv) $>$
18 keV counts (ULD counts) for all layers, and v) $>2$ keV counts
from the veto layer. 
An 8086 microprocessor based system  handles these data and stores them
in 4 Mbits of memory. The data storage is done in different modes
which can be set by commands. The two available modes are 1) count and
spectral mode where the five basic counts are stored in integration time 
of 0.01, 0.1, 1 or 10 s and 64 channel spectra for three layers separately
(top two layers left and right separately) with integration time of
1, 10, 100 or 1000 s. 2) time tagged mode where each event is time
tagged to an accuracy of 0.4 msec (for PPC-3) or 0.8 msec (for PPC-1 and PPC-2)
 and for each event 8 channel linear spectral
information and  layer information are also stored. The data storage can be stopped
and started by the use of time-tagged commands.

The IXAE instrument is a part of the Indian Remote Sensing satellite IRS-P3,
which also includes a remote sensing camera and an oceanographic 
instrument.  IRS-P3 was launched using the Polar Satellite Launch Vehicle
(PSLV) on 1996 March 21 from Shriharikota Range, India. The satellite is in 
a circular orbit at an altitude of 830 km and inclination of 98$^{\circ}$.
Stellar pointing for any given source is done by inertial pointing 
by using  a star tracker. The pointing accuracy is about 0.1$^{\circ}$. 
The observing
time for the 3-axes stabilized stellar pointing mode is available for 3 to 
4 months in a year. The high inclination and high altitude orbit is found to be 
very background prone and the useful observation time is limited to
the latitude ranges typically from -30$^{\circ}$ S to +50$^{\circ}$ N.
Further, the large extent of the South Atlantic Anomaly (SAA) region
restricts the observation to about 5 of the 14 orbits per day. The 
observations are made in the selected latitude regions by using
time-tagged commands, which either reduce the high voltage to the 
non-operating region and stop data acquisition. The data is down loaded
typically twice per day.

The PPCs were first switched-on on 1996 April 30 and Cygnus X-1 was observed
in its hard state between April 30 and May 9. The source  was again
observed, in its soft state, between July 4 and July 10. The log 
of observation is given in Table 1, for only those observations which
are used for the present analysis. The observed count rates as normalized to
the Crab flux from a calibration study undertaken in 1996 December, and the
 binary phase of Cyg X-1 calculated from the ephemeris given by Gies \&  
Bolton (1982), are also given in the table.

\subsection{Background modeling    }

 All the three PPCs are co-aligned and hence there are no simultaneous
background measurements. For each set of observation for a given source
(lasting for a few days), background count rates are measured   before 
and after the source observation by pointing the PPCs at regions in the
sky close to the source (about 5$^{\circ}$), but free of any known 
X-ray sources. As mentioned earlier, the IRS P-3 satellite is in a 
high altitude and high inclination orbit, which results in very high
charged particle induced background count rates in high latitude regions
and South Atlantic Anomaly region.  The good observing regions are generally
restricted to latitude ranges from -30$^{\circ}$ S to +50$^{\circ}$ N.

It is known that the particle background in space environment tracks 
well with the McIlwain's L parameter and the Earth's magnetic field, B
(McIlwain 1977).  For the present observations, 
it is found that the background rates are relatively stable at low 
magnetic filed values ($<$ 0.4 G) and low McIlwain's L parameter ($<$
1.2). To further quantify the background value, we tried to correlate
the observed background rates in the various channels with the particle 
indicators like L, B, ULD count rate and veto count rate. We find that 
for B $<$ 0.4 G and L $<$ 1.2 the observed count rates are well correlated
with the ULD count rates. In Figure 1 we show  the background count
rates obtained in PPC-2 top  layer, plotted against the ULD count rates. The
integration time for each data point is 100 s. We find a correlation
between the two with a correlation coefficient of 0.98. The relationship
between the two quantities can be described by a linear relation and the
best fit straight line is also shown in the figure. 
Similar linear relations are found between the observed background counts
in different layers of all the PPCs and the following prescription is adopted
for background modeling: i) take data only when B $<$ 0.4 G and L $<$ 1.2; 
ii) establish a linear relationship between the background counts and the 
ULD counts and iii) use this relationship for predicting the background at other
times (see Agrawal et al. 1997, for details).

The background subtracted count rates for each of the PPCs in each
observing period of the satellite's orbit are listed in Table 1, after
normalizing them to the observed count rates from Crab calibration. The average
source flux increased by about a factor of 2 from the hard state to the 
soft state. 
 The background subtracted count rates in 1 s time resolution is 
given in Figure 2  for two of the typical observations in the hard state
(top panel) and the soft state (bottom panel) of the source.

\section{Analysis and results             }

\subsection{Power density spectrum }

The light curves show dramatic differences in the two spectral
states when seen in 1 s time resolution (see Figure 2). In the
soft state there are large intensity variations at seconds to minutes
times scales.  To 
quantify these differences we have obtained the power density spectrum
(PDS) for the two states. PDS are obtained for individual data segments
and then co-added, using the XRONOS software package. The PDS are
normalized to the squared fractional rms per unit frequency.
In Figure 3, the observed PDS is shown for the hard state (1996 May). The
data were acquired in 1 s mode as well as 0.4 ms mode and the PDS were 
generated separately for each of the observations and co-added. 
As can be seen from the figure, the PDS at low frequencies
(0.01 Hz to 0.3 Hz) is very flat with a power-law index of -0.09. At higher
frequencies, the PDS steepens and it has a power-law index
of -1.1, with a break frequency of 0.23 Hz.  
The rms variability at the break frequency is 26\%.
There are indications
of additional structures at higher frequencies. A single power-law for
frequencies above 0.23 Hz gives a $\chi^2$ of 197 for 65 degrees of 
freedom and the $\chi^2$ improves to 109 if we allow for one
more steepening at higher frequencies.
The fitted parameters are: additional 
break frequency at 2.5 Hz, power-law index between 0.23 and 2.5 Hz is -1.0,
power-law index between 2.5 and 10 Hz is -1.9.
In Figure 4, the observed PDS for the 1996 July observations (the soft state)
 are shown. 
The power-law index for frequencies above 0.03 Hz is -0.39 and
there is no indication of a break in the slope. 

The general trend of the observed PDS is similar to that obtained using 
the PCA data (Belloni et al. 1996; Cui et al. 1997a). In the soft state the PDS is a
simple power-law. In the transition state, the PDS shows a break at around
0.3 $-$ 0.7 Hz, and the power-law index is between $-$0.6 to $-$0.3
below this frequency.
 The PDS obtained by us in the hard state, though has
a similar break frequency, is  much flatter at lower frequencies
(power-law index $-$0.09).
It is interesting to note that the PDS shape obtained
by us in the hard state agrees very well with that traditionally
 obtained in the hard state by other observers. 
Belloni \& Hasinger (1990a) have analyzed about 30 EXOSAT observations
on Cyg X-1 and found that the PDS is very flat below a break frequency,
follows a power law with slope about $-$1 up to about 1 Hz, and then steepens
to a slope of roughly -2. This is very similar to the behavior detected by us.
The break frequency shows a negative correlation with the rms variability
at the break frequency (Belloni \& Hasinger 1990a)
 and the value of the break frequency (0.23 Hz) and the
rms variability (26\%) obtained by us, agrees very well with this 
correlation. It is interesting to note that the PDS in the hard state
has a similar shape at all times, including immediately prior to a state
transition, as observed by us. During the transition period, the 
PDS parameters show a trend of change such that the power at low 
frequencies increases and comes close to that seen in the soft state.

One of the major differences between the PDS in the soft and the hard
states is the power-law index below 0.2 Hz. If the same shape extends 
to still lower frequencies, the variability characteristics should
be substantially different in the two states at time scales of a few days. 
To investigate this, we
have obtained the archival data from the ASM on-board the RXTE, in the
hard and the soft states, centered around our observations. 
We have also obtained the contemporaneous ASM data on Crab, for calibration
purpose. We find that in the hard state, Cyg X-1 has a average count rate
of 35.9 s$^{-1}$ (compared to 75.8 s$^{-1}$ in Crab). The rms variability
is 25\%, compared to 3.1\% for the Crab (which can be taken as due to 
observational errors). In the soft state Cyg X-1
shows 25\% variability (with an average count rate of 76.1 s$^{-1}$) compared
to 3.5\% for the crab (74.4 s$^{-1}$ average count rate). Hence we can conclude
that there is no significant difference in the variability over   
time scales of days in Cyg X-1 in the two spectral states. To compare this 
variability to the higher frequency range, we have obtained the
PDS for the ASM data and shown these also in Figures 3 and 4.
It should be noted here that the ASM data are not contiguous and hence
the PDS may not be an accurate representation, but the large rms value
is reflected in the PDS also. 
The power-law index for frequencies between 10$^{-5}$ Hz and 0.02 Hz 
is estimated to be -0.98 for the soft state and -1.02 for the hard state. 
Since the EXOSAT data (Belloni \& Hasinger 1990a) showed that the hard state
PDS is flat all the way down to 10$^{-3}$ Hz, the very low frequency
variations may be unrelated to the spectral states and may have a common
source mechanisms for the two spectral states.

\subsection{Shot  statistics  }

Traditionally, the time variability in Cyg X-1 was sought to be explained
by models invoking random occurrences of shots. The hard state PDS were
qualitatively explained by Belloni \& Hasinger (1990a) by a shot noise model
with a distribution in shot times.
The observed distribution of shots in Cyg X-1 has also been explained under the
premises of self-organized criticality (SOC) model (Mineshige et al. 1994a). 
In this model it is assumed that the inner portions of accretion disk 
are composed of numerous small reservoirs. If a critical mass density is
reached at some reservoir, an instability gets established.   This model
predicts a power-law distribution of shot energy and duration
(Mineshige et al. 1994b). The observed distribution of shots are, however,
exponential (Negoro et al. 1995), and it was reconciled
with  the SOC model by assuming a gradual mass diffusion in the reservoir.
To investigate whether the shot distribution  obtained by Negoro et al. (1995)
has a similar shape in both the spectral states, we have subjected our data
to a shot distribution analysis. We  have taken the
total energy in the shots and the shot width rather than the shot peak
intensity.  For this purpose, the
light curves are taken with 1 s binning. Background subtracted counts
are compared in each bin with a running average of length 20 times the bin 
width. When the observed
counts in a bin exceed the average value, a shot is deemed to have started.
All the successive bins are considered to be a part of the same shot if
all of them have count rates in excess of the running average. The duration
and the excess counts in each shot are calculated and histograms are 
generated.  The excess counts are normalized to the average count rate
such that the excess counts are expressed in equivalent width in seconds.

The number distribution for the shot equivalent width is shown in Figure 5  
for the hard state observations. The distribution for Poissonian noise
of equivalent average count rate is much lower than the observed 
distribution. We find that the distribution can
be described by an exponential function with an e-folding constant
of 0.35 s. We find that other simple functional forms like power-law
etc do not fit the observations.

A similar shot statistics is shown in Figure 6 for the soft state observations.
The distribution is much steeper with an e-folding constant of 0.2 s.
Apart from this, there are a large number of high intensity
shots over and above that described by the exponential function. 
We have also generated the shot distribution
for the soft state when the count rates are taken in 0.1 s bins (Figure 7). The
distribution is steep with an e-folding constant of 0.04 s. We also
find a shot with large excess counts (with an equivalent width of 0.92 s). 
The formal probability that this
shot occurred at random assuming a shot occurrence rate described by an 
exponential function 
is 2 $\times$ 10$^{-7}$. This shot was detected by PPC-1 on 1996 July 5. 
The light curve for this shot is shown in
Figure 8  for the total counts (top panel), 2-6 keV counts (middle panel)
and 6-18 keV counts (bottom panel). The shot can be described by   
exponential rise and decays with a time constant of 0.4 s. 

The shot width distribution also shows an exponential distribution. We find that
in the hard state it has a time constant of 1.4 s, and the distribution
becomes flatter in the soft state with a time constant of 2.2 s. We also
find that the distribution cannot be described by other simple functional
forms like power-law etc.

\section{Discussion }

Detailed observations of Cyg X-1 made using the PCA on-board the RXTE 
(Cui et al., 1997a,b; Belloni et al., 1996) have shown that the power density
spectrum (PDS) of the source changes smoothly during the spectral transition
from the hard state to the soft state and reverts back when there is a transition
back to the hard state. 
The power-law index for the PDS below 0.3 Hz for the transition state was 
found to be between -0.3  to -0.7.
The observations carried out by the PPCs in 1996 May
pertain to the hard state of the source, where there were no PCA observations.
We find that the index is still flatter in the hard state, about -0.09, and the
PDS shape agrees very well with the traditional hard state PDS obtained
earlier (Belloni \& Hasinger 1990a).
These observations reinforce the conclusion obtained by Cui et al. that
the changes in the PDS smoothly follow the spectral changes.

Chakrabarti  \&  Titarchuk (1995) have worked out a complete solution of viscous
transonic equations and have identified the various X-ray emission
regions in an accretion disk.
Here we attempt to identify the different components in the PDS with the
different regions of the accretion disk as described by Chakrabarti \&
Titarchuk (1985). Most of the white noise component in the PDS can originate
in the post-shock region. In the soft state, the Compton cooling is very
efficient resulting in bulk motions which can lead to much higher red noise
in the PDS. The very low frequency component in the PDS (at $<$0.01 Hz) identified in the
present work appears to be independent of the spectral states. The origin of
this component could be from the pre-shock sub-Keplerian component.

The distribution of shots in Cyg X-1 has been explained under the
premises of self-organized criticality (SOC) model (Mineshige et al. 1994a). 
Negoro et al. (1995) have examined this model by analyzing the 
Ginga data for the hard state of Cyg X-1. They have also put a selection 
criterion for shots, namely, the shot peak count rates should be 
higher than the running average by   a pre-determined factor, p. It
was found that the shot widths follow an exponential distribution with time
constants of 1.8 s, 8.4 s and 23.8 s for values of p 1.5, 2 and 2.35,
respectively. For our analysis, we have taken all shots (p is 1) and
find that the
shot width has an exponential distribution with time constant of 1.4 s. This
value appears to be consistent with 1.8 s (for p $=$ 1.8), since the 
time constant decreases with decreasing p. 
The important conclusion that can be drawn from our work is
that the shot width as well as the total shot intensity has an exponential
distribution in both the spectral states.
The distinct change in the 
shot distribution for the two spectral states is a new input for the
shot modeling and it probably indicates the differences in the basic
model parameters like the critical mass density, diffusion coefficient etc. 

The large shot detected in our work appears similar to the shape of
the co-added shots obtained from the Ginga data (Negoro et al. 1994).
They have found two time constants each for the rise and decay phase
of values 0.1 and 1 s, respectively. The value obtained in the present work
(0.4 s) lies in between these two values.

\section{Conclusions}

We have analyzed the data obtained from the Indian X-ray Astronomy
Experiment (IXAE) on Cyg X-1 in its two spectral states. A method
is evolved for background subtraction.  The timing analysis has shown
that:

1. The power density spectrum (PDS) shows substantial differences
in the two spectral states: the power-law index between 0.03 to 0.3
Hz changed from -0.09 to -1.02 when the source changed from the hard
to the soft state.

2. A new component in the PDS below 0.03 Hz has been identified and
this component appears to be independent of the spectral state.

3. The shot statistics like the duration and the energy in shots
show exponential distribution and the time constant shows significant
difference in the two spectral states.

4. A large individual shot has been identified and it has an exponential 
rise and decay with a time constant of 0.4 s.

\begin{acknowledgements}

It is a pleasure to acknowledge the support given by Shri K. 
Tyagarajan, Project Director, IRS-P3, Shri R.N. Tyagi, Head, IRS PMO,
Shri R. Aravamudan, Director, ISAC, Mission Planning and Operation Group
of ISRO, Dr. T.K. Alex and his team at LEOS who designed and operated the
star tracker. We also thank all the technical and
engineering staff of ISAC in making the IXAE project a success. We are
extremely grateful to the engineers, scientific and technical staff
of the X-ray and gamma-ray Astronomy group of TIFR, Technical Physics Division
of ISAC and the ISRO Satellite Tracking Center. Thanks are also due to
the RXTE team for making the ASM data public.

\end{acknowledgements}

\newpage

\begin{table*}
\caption{Log of observation}
\begin{tabular}{|ccc|lll|lll|}
\hline
&~~&&&&&&&\\
 Observation &Cyg X-1 & Time       & Live & time & (sec) & Observed & flux &(mCrab)  \\
 date        &binary  & resolution & PPC-1& PPC-2& PPC-3 & PPC-1 & PPC-2 & PPC-3 \\
 time        &phase   & (sec)      &    &        &       &       &       &       \\
&~~&&&&&&&\\
\hline
\hline
&~~&&&&&&&\\
   Apr 30  12:10 &    .44 & 1.0  &  -- &  895 &  697 &   -- &  503 &  514 \\ 
   Apr 30  13:50 &    .45 & 1.0  &   -- &  898 &  897 &   -- &  434 &  437 \\ 
   May  1  11:48 &    .61 & 1.0  &   -- &   -- &  696 &   -- &   -- &  633 \\ 
   May  1  13:30 &    .63 & 1.0  &   -- &   -- &  800 &   -- &   -- &  562 \\ 
   May  1  15:10 &    .64 & 1.0  &   -- &   -- &  896 &   -- &   -- &  573 \\ 
   May  3  11:05 &    .97 & 1.0  &  799 &  800 &   -- &  431 &  417 &   -- \\ 
   May  3  12:46 &    .98 & 1.0  &  795 &  899 &   -- &  395 &  384 &   -- \\ 
   May  3  14:28 &    .99 & 1.0  &  696 &  699 &   -- &  398 &  383 &   -- \\ 
   May  3  23:43 &    .06 & 1.0  &  694 &   -- &  699 &  437 &   -- &  471 \\ 
   May  4  04:58 &    .10 & time-tagged  &  338 &   -- &  338 &  396 &   -- &  420 \\ 
   May  9  06:28 &    .00 & 1.0  &  477 &  675 &  500 &  319 &  308 &  356 \\ 
   May  9  14:03 &    .06 & 1.0  &   -- &  900 &  898 &   -- &  437 &  445 \\ 
   May  9  23:18 &    .13 & 1.0  &  700 &  698 &  697 &  390 &  384 &  390 \\ 
   May 10  00:58 &    .14 & 1.0  &  800 &  797 &  800 &  365 &  354 &  365 \\ 
   May 10  02:38 &    .15 & 1.0  &  899 &  896 &  900 &  364 &  346 &  379 \\ 
   May 10  06:03 &    .18 & 1.0  &  695 &   -- &  799 &  362 &   -- &  363 \\ 
   May 10  13:45 &    .24 & 1.0  &  628 &   -- &  644 &  449 &   -- &  439 \\ 
   May 10  15:23 &    .25 & 1.0  &  499 &   -- &  499 &  430 &   -- &  422 \\ 
   May 10  17:09 &    .27 & time-tagged  & 1085 &   -- & 1085 &  308 &   -- &  298 \\ 
   May 10  22:58 &    .30 & 1.0  &  599 &  601 &  597 &  418 &  386 &  440 \\ 
   May 11  00:36 &    .32 & 1.0  &  799 &  797 &  796 &  430 &  385 &  455 \\ 
   May 11  02:16 &    .33 & 1.0  &  897 &  995 &   -- &  337 &  283 &   -- \\ 
&~~&&&&&&&\\
   Jul  5  14:21 &    .24 & 0.1  &  700 &   -- &   -- &  756 &   -- &   -- \\ 
   Jul  8  13:16 &    .77 & 0.1  &  800 &  900 &   -- &  773 &  857 &   -- \\ 
   Jul  8  15:07 &    .79 & 0.1  &  251 &  335 &   -- &  939 & 1079 &   -- \\ 
&~~&&&&&&&\\
\hline
\end{tabular}
\end{table*}

\newpage

\begin{figure}
\vskip 6.5cm
\includegraphics{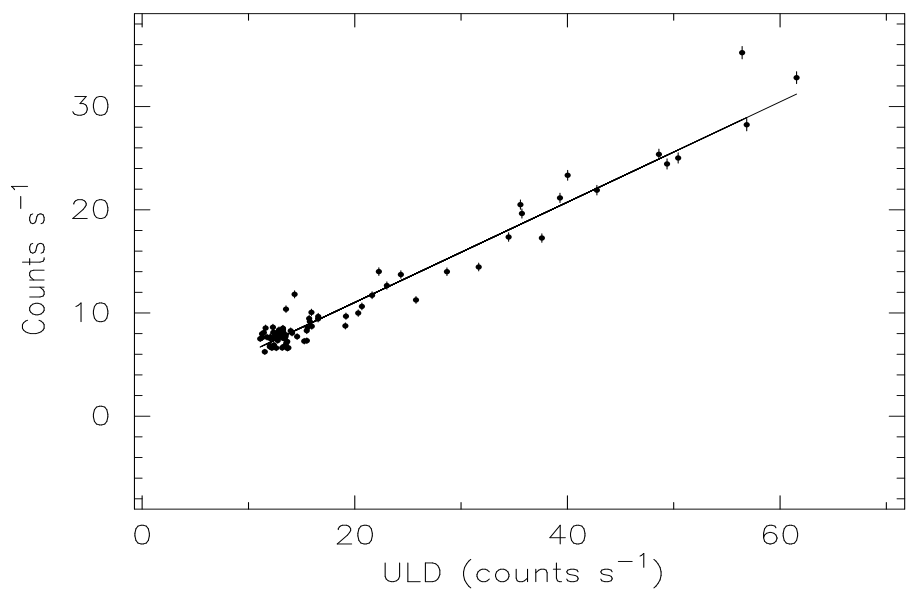}
Š\caption[]{The observed background counts in PPC2r top layer
are plotted against the ULD (upper level discriminator) counts, which is
a measure of the particle background. The integration time for the data points
is 100 s.  The continuous line in the figure
is a least square straight line fit. 
}
\end{figure}

\begin{figure}
\vskip 6.5cm
\includegraphics{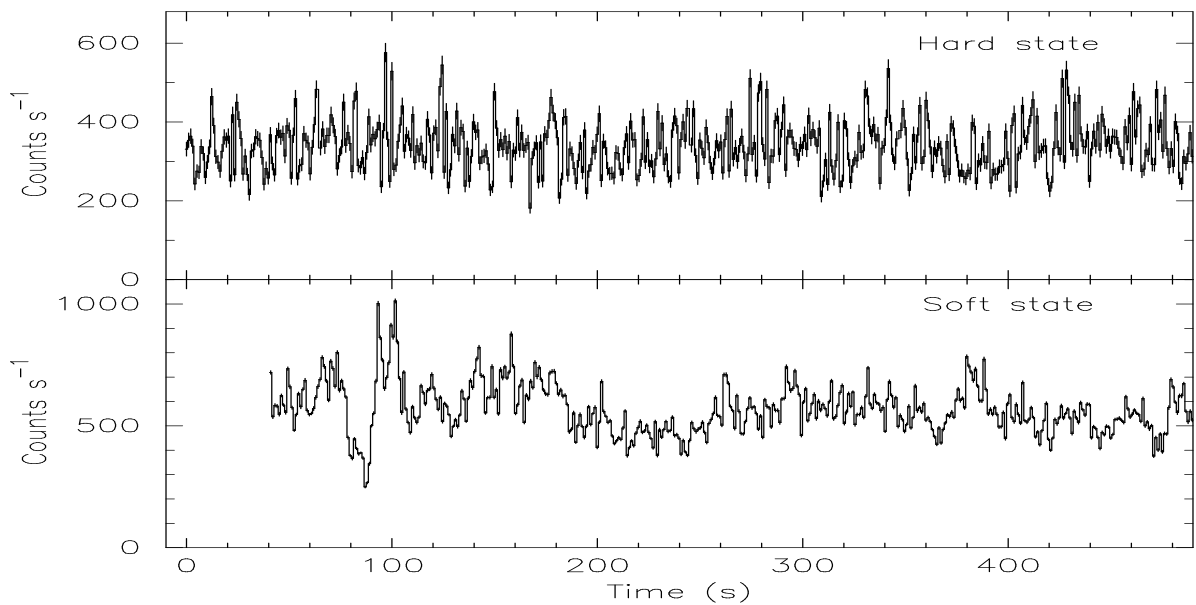}
\caption[]
{ The background  subtracted count rates from PPC-2 obtained during the 
hard state of Cyg X-1 (1996 May) are  shown in the top panel. Count rates from the
same detector obtained in 1996 July (during the soft state of Cyg X-1)
 are shown in the
bottom panel.  The time resolution is 1 s. The start time is
arbitrary. 
}
\end{figure}

\begin{figure}
\vskip 6.5cm
\includegraphics{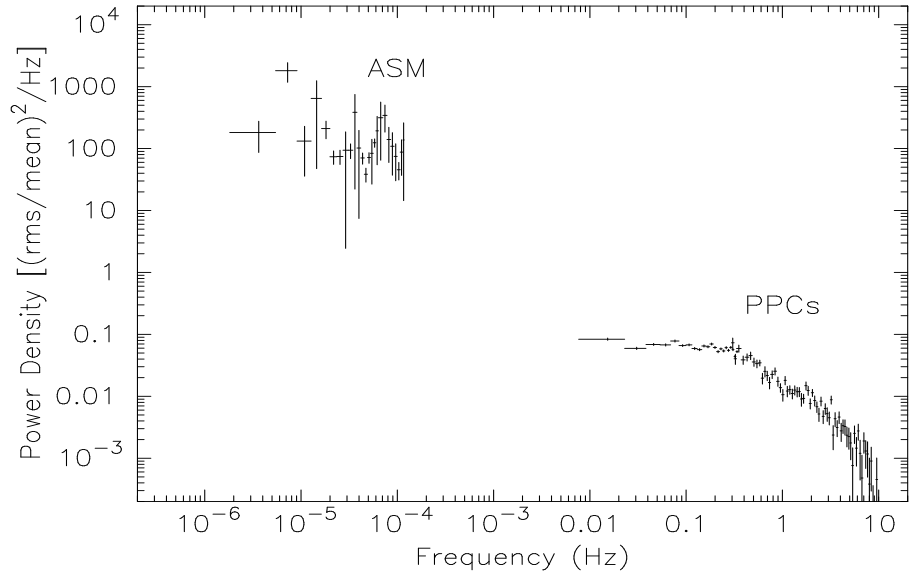}
Š\caption[]
{ The power density spectrum (PDS) for the hard state of Cyg X-1 obtained
during 1996 May from the PPC observations as well as from the ASM
data. }
\end{figure}

\begin{figure}
\vskip 6.5cm
\includegraphics{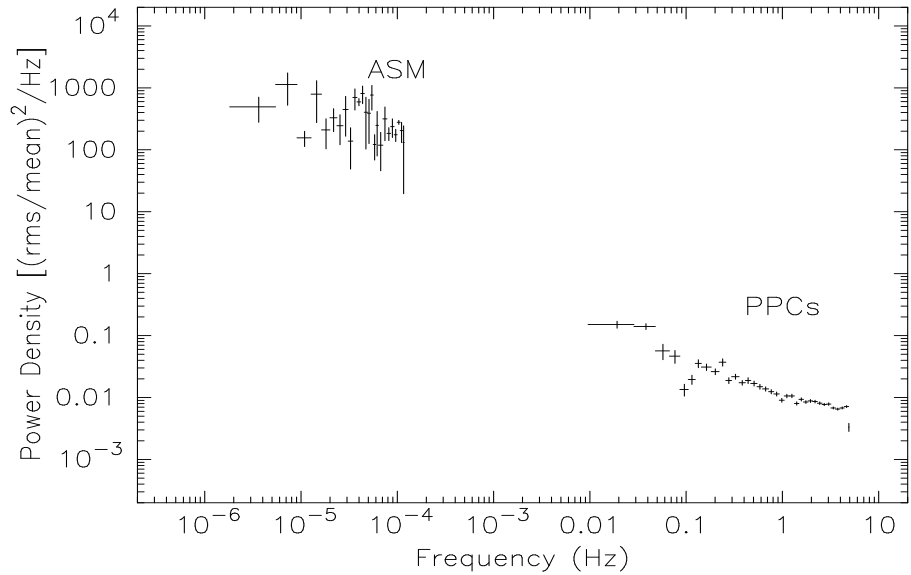}
Š\caption[]
{ Same as figure 3, but the data are for the 1996 July (soft state) observations. }
\end{figure}

\begin{figure}
\vskip 6.5cm
\includegraphics{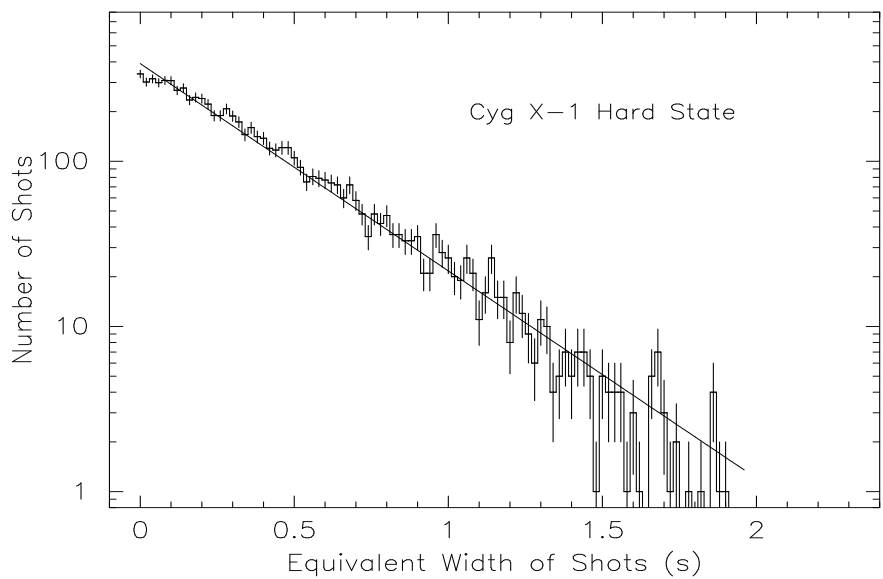}
Š\caption[]
{ The Shot  statistics of Cyg X-1 for the hard state (1996 May) observations.
The total number of shots  of a given total excess counts (normalized 
to the local mean)
are plotted against the excess counts (differential plot).
The units of the excess counts are equivalent width in seconds.
The best fit exponential function is shown as a straight line, and the 
derived time constant is 0.35 s.
Time resolution of the light curve used for generating shot  statistics is
1s. }
\end{figure}

\begin{figure}
\vskip 6.5cm
\includegraphics{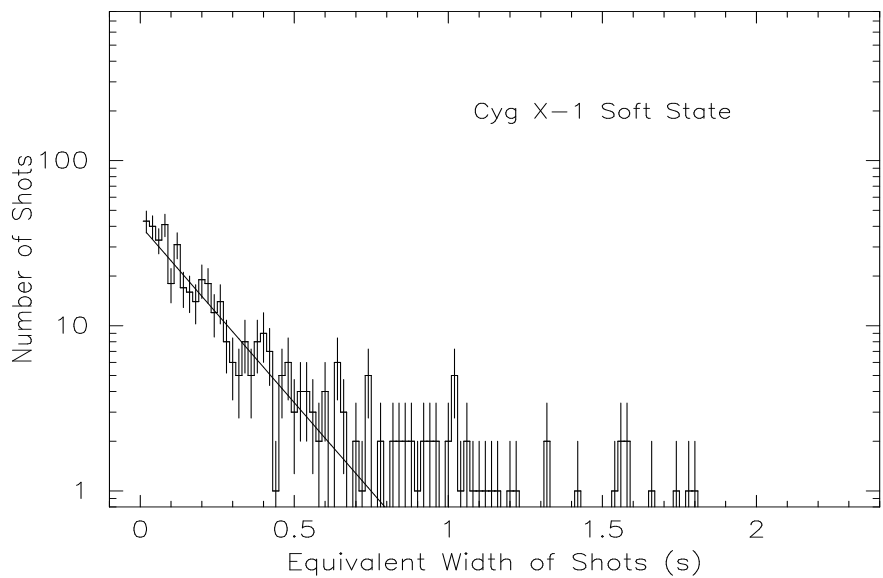}
Š\caption[]
{ Same as figure  5, but the data are for the 1996 July (soft state)
observations. The exponential fit has a time constant of 0.2 s. }
\end{figure}

\begin{figure}
\vskip 6.5cm
\includegraphics{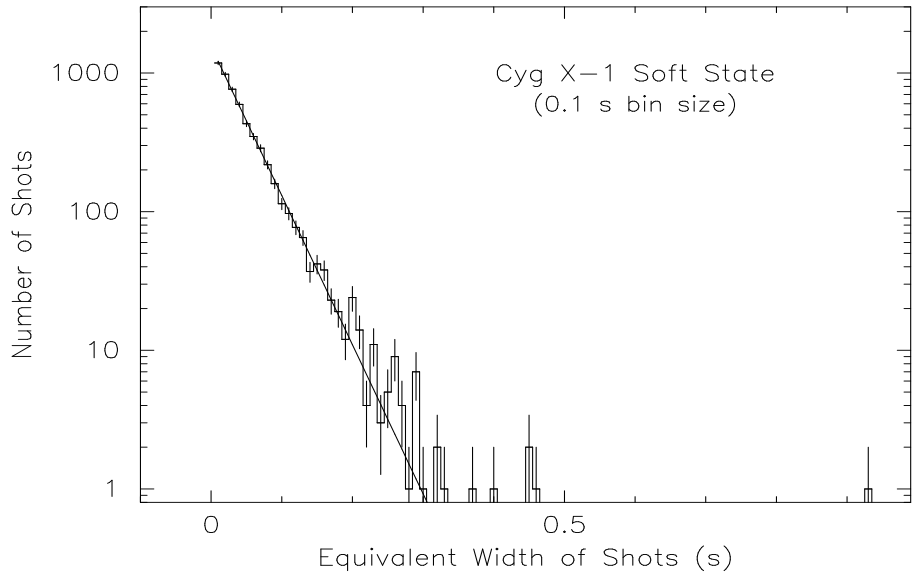}
Š\caption[]
{ Same as figure  6, but the time resolution for the flare statistics
generation is 0.1 s. The exponential fit has a time constant of 0.04 s. }
\end{figure}

\begin{figure}
\vskip 6.5cm
\includegraphics{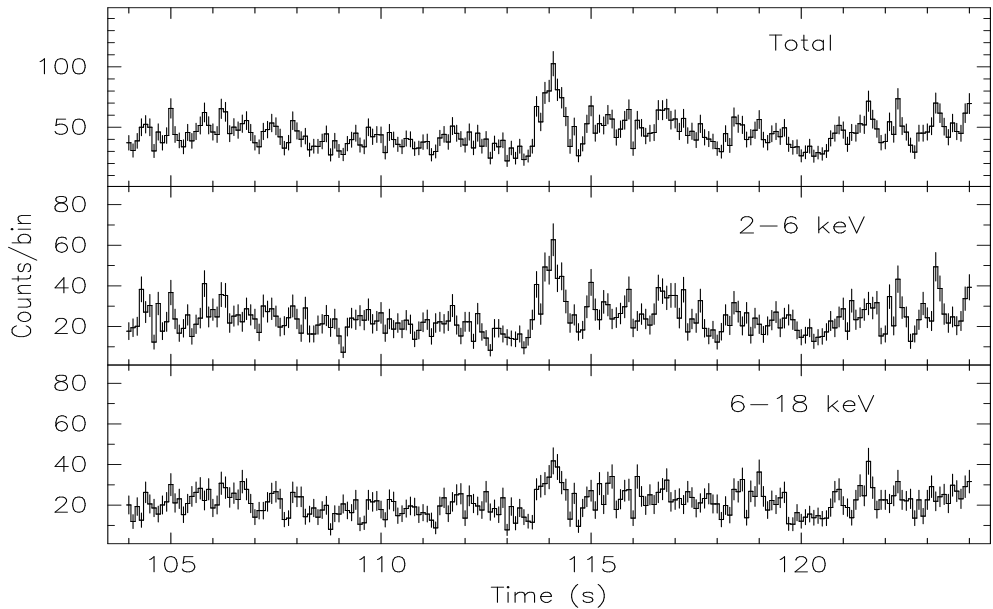}
Š\caption[]
{ The details of the largest shot  observed during the soft state (1996 July).
The observed counts per bin (for a bin size of 0.1 s) are shown for
PPC-1 total counts (top panel), 2 keV $-$ 6 keV counts (middle panel) and
6 keV $-$ 18 keV counts (bottom panel). The time is given in seconds starting from
1996 July 5 14:21 UT. }
\end{figure}

\end{document}